\documentclass[conference]{IEEEtran}
\IEEEoverridecommandlockouts
\usepackage[utf8]{inputenc}
\usepackage[T1]{fontenc}
\usepackage{cite}
\usepackage{amsmath,amssymb,amsfonts}
\usepackage{graphicx}
\usepackage{booktabs}
\usepackage{array}
\usepackage{url}
\usepackage{textcomp}
\usepackage{xcolor}
\usepackage{orcidlink}
\usepackage{marvosym}

\begin{document}
\title{Multi-View Fusion for Encrypted C2 Detection: A Leakage-Controlled Measurement Study of Evaluation Pitfalls}

\author{\IEEEauthorblockN{Hoang-Huy Nguyen-Huu\textsuperscript{1}, Van-Tri Phan\textsuperscript{1,(\Letter)}, and Khuong Nguyen-An\textsuperscript{2,3,(\Letter)}\,\orcidlink{0000-0002-9910-6387}}
\IEEEauthorblockA{\textsuperscript{1}Academy of Cryptography Techniques, Ho Chi Minh City Campus, Vietnam\\
\textsuperscript{2}Faculty of Computer Science and Engineering,\\ Ho Chi Minh City University of Technology (HCMUT),\\ 268 Ly Thuong Kiet Street, Dien Hong Ward, Ho Chi Minh City, Vietnam\\
\textsuperscript{3} Vietnam National University Ho Chi Minh City, Linh Xuan Ward, Ho Chi Minh City, Vietnam\\
Emails: \texttt{huylhp1731999@gmail.com}, \texttt{phanvantri@actvn.edu.vn}, \texttt{nakhuong@hcmut.edu.vn}}}

\maketitle
\begin{abstract}
Command-and-control (C2) traffic increasingly hides within TLS, so defenders now apply machine learning to traffic metadata. Many studies assume that combining two metadata views, namely flow statistics and TLS handshake fingerprints, improves both accuracy and robustness. We tested this assumption on 17,577 TLS flows from 62 real Cobalt Strike captures. Our evaluation removes the data leakage that leads to overly optimistic reported scores. We report three findings that matter more than the fusion result itself. First, an incorrect preprocessing step increases the F1 score by 0.28. This step computes the frequency encoding across the entire dataset rather than within each cross-validation fold. The increase is about ten times larger than any real effect we measured. Second, both the labels and the behavioral features depend on the destination address. Because of this, the 17,577 flows form only 2,132 independent groups, and the positive rate of 55.1\%, which looks balanced, drops to 4.2\%. Therefore, class balance is just a result of how we analyze the data, specifically whether we count flows or endpoints, and not a real feature of the task. Third, 20 of the 62 captures (32\%) have no TLS flows to any known C2 address, so they contain only benign samples. We checked these captures directly and confirmed that this is a gap in the ground truth, not a labeling error. In this context, fusion beats the best single view by only 0.022 in F1. When an attacker forges both feature surfaces simultaneously, every model performs worse than a simple baseline that always predicts positive (F1 = 0.711). For encrypted C2 detection, the evaluation design is not a preliminary step. It \emph{is} the main result.
\end{abstract}

\section{Introduction}

Command and control (C2) is a central part of an intrusion. After an attacker opens a stable channel to a compromised host, every later stage depends on it, including lateral movement, privilege escalation, and data exfiltration. For this reason, MITRE ATT\&CK treats the C2 tactic (TA0011) as an important detection target \cite{ref1}. Most C2 channels now run over TLS, placing the problem between two techniques: Web Protocols (T1071.001) and Encrypted Channel with Asymmetric Cryptography (T1573.002). At the packet level, a C2 session over HTTPS with a valid certificate looks like normal web traffic, and signature-based detection cannot read the encrypted payload \cite{ref2}.

A common response is to use traffic metadata instead of the payload. Anderson and McGrew showed that flow data, such as packet sizes, the time between packets, and flow durations, still carries a useful signal even when the payload is unreadable \cite{ref3}, \cite{ref4}. The visible part of the TLS handshake gives a second source of information. It includes the negotiated version, the cipher suites, the server name (SNI), the certificate fields, and the JA3 and JA3S fingerprints \cite{ref5}. These two sources describe different aspects of the same session, so combining them seems appealing. Today, this combination, called multi-view fusion, is a common approach in the literature.

Two problems make this evidence difficult to trust. The first problem is the evaluation method. Arp et al.~showed that many security ML results report F1 above 0.99 but drop once data leakage is removed \cite{ref6}. Encrypted C2 detection is especially exposed to this issue. Flows from the same capture share TLS fingerprints, server identities, and beacon intervals. If we split the data by individual flow, the model can memorize features from one campaign rather than learning behavior that generalizes to new data. The second problem is that fusion is usually tested only under clean conditions, which assume that the test data follows the same distribution as the training data. An attacker is exactly the person who breaks that assumption.

This paper studies both problems together. It shifts the question from \emph{how well fusion detects} to \emph{what the evaluation itself reveals}. In our study, the largest effects are measurement artifacts, not real gains in detection. Measuring these artifacts is more useful than reporting one more fusion result.\\

\noindent \textbf{Contributions.} The first three contributions of this work are measurement findings; the fourth is an evaluation. (i)~Fitting the frequency encoding globally rather than inside each fold raises the TLS-view F1 by 0.28, larger than any real difference in the study.
(ii)~Class balance is an artifact of the unit of analysis: the labels and nine beacon features all depend on the destination, so the 17,577 flows reduce to 2,132 groups, and the positive
rate falls from 55.1\% to 4.2\%. (iii)~The ground truth is incomplete: in 20 of the 62 captures, no TLS flow reaches any published C2 address, which is a limit of threat intelligence, rather than a
labeling error, and we bound a related label-noise source at 11.2\% of flows. (iv)~Fusion outperforms the best single view by only
0.022 F1, and under attack on both surfaces, every model falls below a trivial baseline, so clean-condition ranking does not predict adversarial ranking.

\section{Related Work}

Classifying TLS traffic without reading the payload relies on one fact: packet sizes and timing still carry a signal after encryption. Anderson and McGrew first showed this with contextual flow data, then extended it to malware traffic without decryption \cite{ref3, ref4}. Later work built open frameworks for reproducible encrypted and
malicious traffic classification \cite{ref7} and applied transformer models that read packet bytes directly \cite{ref8}. A second source of information is the plaintext part of the TLS handshake. JA3 and JA3S summarize the ClientHello and the ServerHello as short fingerprints \cite{ref5}. This source is weak against an attacker because a malleable C2 profile can send a ClientHello that looks like a browser, and the fingerprints also vary across software versions, which led to the newer JA4+ scheme \cite{ref9}. A third source is periodic callback timing, used, for example, by BAYWATCH \cite{ref10}. Studies combine these sources by early or late fusion. Stacked generalization \cite{ref11} is the learned form, and the gain from any ensemble is limited by how differently its models make errors \cite{ref12}.

A weakness runs through this whole line of work: the evaluation design. Many studies split the data by individual flow at random. This allows flows from one capture to appear in both the training and test sets, which is exactly the problem our protocol removes. Several papers document why this matters: Sommer and Paxson show poor transfer from the laboratory to deployment \cite{ref2}; Arp et al.~show that many scores fall once leakage is removed \cite{ref6}; TESSERACT shows bias across time and data sources \cite{ref13}; and Kapoor and Narayanan find leakage in 294 papers across 17 fields \cite{ref14}. On the attack side, Pierazzi et al.~separate feature-space attacks from problem-space attacks \cite{ref15}, and our evasion is a feature-space attack. Against this background, our contribution is to treat three common gaps as measurements instead of assumptions: leakage
is usually assumed rather than measured; fusion is reported as an outcome rather than explained as a mechanism; and robustness, when tested at all, is tested on only one surface, which biases the
conclusion toward whichever surface was left untouched.

\section{Threat Model}

We consider an attacker who runs a Cobalt Strike team server. The attacker wants to keep the C2 channel operational while avoiding detection by a deployed metadata classifier. The attacker can change two groups of features at different costs.\\

\noindent \textbf{TLS surface (low cost).} A malleable C2 profile changes the ClientHello and therefore the JA3 fingerprint. The server configuration changes JA3S. A free ACME certificate replaces a self-signed one. All of these are edits to configuration files and take only minutes.\\

\noindent \textbf{Flow-behavior surface (higher cost).} Increasing the \texttt{sleep} and \texttt{jitter} settings removes the regular timing. The \texttt{data\_jitter} setting and extra padding blur the packet-size distribution. Cover traffic changes the volume statistics. These changes cost latency, bandwidth, and stability, but none of them is impossible. We therefore reject the common assumption that flow behavior cannot be forged.\\

\noindent \textbf{How we simulate the attack and its limits.} For each C2 flow in the test fold, we replace the features on the attacked surface with values taken from a random benign flow in the same fold. We train the models on clean data and test them on the modified data. This measures robustness at test time; it is not adversarial training. We sweep the fraction of modified flows over $\{0, 25, 50, 75, 100\}\%$. This is a feature-space attack in the sense of Pierazzi et al.~\cite{ref15}. Replacing one feature at a time can produce a value combination with no real network counterpart, and it assumes perfect mimicry, so it gives an upper bound on the attacker's power. It does not model an adaptive attacker who knows the model, nor functional limits. We exclude structural features, namely TCP flag counts and durations, from every attacked surface because changing them requires action at the transport layer.

\section{Data and Methodology}

\subsection{Dataset and Ground Truth}
We collected every malware-traffic-analysis.net (MTA) post whose title mentions Cobalt Strike from 2022 to 2024. This resulted in 63 capture sessions (64 PCAP files, about 1.6 GB), of which we processed 62 into flows.\footnote{We skipped one session (cs\_20230118, 64 flows, 0.4\%) due to a file-traversal bug. Re-extraction confirmed correct labels; we kept 62 sessions to preserve the cross-validation split.} The captures are not spread evenly over time: 74.2\% are from 2022, 24.2\% from 2023, and 1.6\% from 2024. We preferred real incident captures because synthetic ones lack background noise, varied infrastructure, and real user activity. After extraction and view fusion, the corpus has 17,577 TLS flows: 9,688 are C2 (55.1\%) and 7,889 are benign (44.9\%). They span ten loader and dropper families.\\

\noindent \textbf{Labeling.} We label a flow as C2 if and only if its destination address is in the 160-address indicator set published with the matching MTA post. No feature used for classification takes part in labeling, so there is no circularity at the labeling stage. This rule has a known weakness: a backup C2 or exfiltration address that is not in the published list is labeled as benign. We could confirm only 83 of the 160 indicators (51.9\%) via observed TLS handshakes in the captures. We keep the rest based on the reports' authority.\\

\noindent \textbf{Incomplete ground truth.} When we checked the indicator set against the observed flows, we found that 20 of the 62 captures contain no TLS flows to any indicator address. These captures, therefore, contribute only benign samples. We return to this in Section V-D, where we show that it stems from incomplete source data rather than a labeling error.

\subsection{Feature Views}
\noindent \textbf{View A: flow statistics and beacon behavior (67 features).} NFStream 6.6 \cite{ref16} groups packets into bidirectional flows based on the standard 5-tuple with an idle timeout 120 s, active timeout 1800 s. We kept all 57 statistical descriptors without manual filtering. They cover durations, volumes, packet-size moments, inter-arrival-time moments, and TCP flag counts. We left the choice of relevant features to Boruta. We dropped the nDPI application-identification fields to keep the views disjoint. We kept the destination port.

We computed nine beacon descriptors for each (source, destination, destination-port) triple within each capture, from observed timestamps only: the flow count, five inter-arrival-time moments, the FFT peak
power and period, and the idle ratio. We assigned these values to every flow in the triple because they describe an endpoint pair rather than a single flow. A triple with only one flow receives zeros. This keying matters because both the label and the beacon features come from the destination. It is the basis of the finding on the unit of analysis in Section V-A.\\

\noindent \textbf{View B: TLS handshake metadata (16 features).} We extracted these with tshark 4.6: JA3, JA3S, TLS version, cipher and extension counts, SNI, SNI entropy and length, ALPN presence, HTTP/2 negotiation, session ID and ticket presence, certificate chain length, validity period, self-signed flag, and the destination port.

NFStream orients a flow based on the host that sends the SYN, while tshark reports the true source, meaning the ClientHello and ServerHello of a connection appear with reversed 5-tuples. Joining on the raw tuple matched only 52.8\% of flows. Sorting the two endpoints into an undirected key raised the match rate to 100\% and JA3S coverage to 99.7\%.\\

\noindent \textbf{View C: early fusion (82 features).} We concatenated the columns
of View A and View B and removed the duplicated destination port. All three views cover the same 17,577 flows, with the same labels and the
same group assignments. Therefore, any difference in performance comes solely from the feature set.

\subsection{A Protocol That Controls Leakage}

We evaluated using StratifiedGroupKFold with 5 folds, with source capture as the group. Every flow from one capture therefore went entirely into training or entirely into test. We used five folds rather
than ten because splitting 62 groups into ten parts would leave single-capture families such as Gozi missing from some test folds. All configurations shared one fold assignment, so every comparison is paired by construction.

We fitted three transforms on the training fold only and applied them to the test fold: frequency encoding of JA3, JA3S, and SNI (unseen values mapped to zero); median imputation; and Boruta feature selection \cite{ref17}, run separately with a Random Forest and an XGBoost estimator, keeping the intersection.

To check that the protocol prevents leakage, we ran a controlled violation by fitting the frequency encoding globally before the fold assignment. F1 on View B rose by \textbf{0.28}. The reason is direct.
Some JA3S hashes appear almost only in C2 flows, so a global frequency encodes label information that is not available on new captures. This single result motivated the whole study because it is larger than any real effect we measured later.\\

\noindent \textbf{Models.} We used Random Forest \cite{ref18} (300 trees, balanced class weights) and XGBoost \cite{ref19} (300 rounds, learning rate 0.1, logistic loss). We fixed the random seed to 42 throughout. We selected a tree depth with the one-standard-error rule \cite{ref20}, searching $\{3, 5, 7, 9, 12, 16, \text{None}\}$ for Random Forest and $\{2, 3, 4, 5, 6, 8, 10\}$ for XGBoost. We did not use deep learning: tree ensembles remain competitive on mixed tabular data at this sample size \cite{ref21} and expose feature attributions directly, which the robustness analysis needs.

We note and return to in Section VII, that we selected the depth on the same five outer folds used for reporting. The one-standard-error rule reduces the resulting optimism but does not remove it.

\subsection{Fusion Strategies and Diagnostics}

We evaluated three fusion families. \emph{Early fusion} trains one model on the concatenated View C. \emph{Late averaging} takes the mean of the out-of-fold probabilities from four base models at threshold 0.5. \emph{Stacking} fits a logistic-regression meta-learner on the
out-of-fold base probabilities, in a leave-fold-out manner. Six more
decision-level combiners serve as diagnostics.

For the complementarity analysis, an \emph{oracle} combiner is correct whenever at least one base model is correct; it upper-bounds any combiner over the same base models. We also compute four diversity measures on the pooled out-of-fold predictions \cite{ref12}.

\subsection{Statistical Protocol}

We keep the description separate from the inference. We compare configurations across folds with the Wilcoxon signed-rank test at $\alpha = 0.05$. With $n = 5$ paired observations, the smallest possible $p$-value is $2/2^{5} = 0.0625$. The test, therefore, \emph{cannot} reach significance at 0.05, whatever the
effect size. We report the Wilcoxon results as descriptions, writing
``observed improvement'' rather than ``significant improvement,'' and we also do not claim equivalence. For the evasion analysis, we use a cluster bootstrap over the 62 captures (2,000 resamples). It respects the grouping of flows inside a capture and is not limited by the number of folds.

\subsection{Reproducibility}

The experiments ran on Kali Linux (ARM64) with fixed random seeds.
Frequency encoding, imputation, and Boruta all ran inside the
cross-validation loop, with no global preprocessing before the fold
assignment. The library versions are pinned in the released environment. Code and derived data are available at \url{https://github.com/HuyLHP/encrypted-c2-detection.git}.

\section{Results}

We first present clean-condition detection together with three measurement findings: the unit of analysis, complementarity, and incomplete ground truth. We then present the fusion and evasion results that these findings put in context.

\subsection{Clean-Condition Detection and the Unit of Analysis}

\begin{table*}[!t]\centering\caption{Clean-Condition Performance (mean $\pm$ SD over 5 folds)}\label{tab:clean}\small
\begin{tabular}{@{}llccccc@{}}\toprule
\textbf{View} & \textbf{Model} & \textbf{F1} & \textbf{Precision} & \textbf{Recall} & \textbf{ROC-AUC} & \textbf{PR-AUC} \\ \midrule
A (flow+beacon) & RF & \textbf{0.856 $\pm$ 0.131} & 0.850 & 0.882 & 0.877 & \textbf{0.915} \\
A & XGB & 0.775 $\pm$ 0.130 & 0.821 & 0.776 & 0.801 & 0.846 \\
B (TLS) & RF & 0.762 $\pm$ 0.123 & 0.848 & 0.742 & 0.849 & 0.881 \\
B & XGB & 0.829 $\pm$ 0.183 & 0.810 & 0.857 & 0.869 & 0.906 \\
C (early fusion) & RF & 0.859 $\pm$ 0.131 & 0.858 & 0.881 & 0.868 & 0.908 \\
C & XGB & 0.797 $\pm$ 0.140 & 0.808 & 0.823 & 0.815 & 0.855 \\
Late averaging & --- & \textbf{0.878 $\pm$ 0.140} & 0.856 & 0.918 & \textbf{0.883} & 0.911 \\
Stacking & --- & 0.781 $\pm$ 0.123 & 0.807 & 0.792 & 0.846 & 0.888 \\
\textbf{Always-positive} & --- & 0.711 $\pm$ 0.030 & 0.551 & 1.000 & 0.500 & 0.551 \\
\bottomrule\end{tabular}\end{table*}

Late averaging reaches the highest F1 (0.878, Table~\ref{tab:clean}).
Early fusion with Random Forest reaches 0.859, only 0.003 above the best single view and well inside the fold-to-fold standard deviation of
0.131. We included the always-positive baseline (F1 = 0.711) in the results table itself. The true margin of the strongest model over a trivial classifier is only 0.145, not the 0.856 that the raw number suggests.\\

\noindent \textbf{Per-fold structure.} The average hides a two-part pattern. The five folds of A-RF are 0.946, 0.604, 0.872, 0.965, and 0.893, and the standard deviation of 0.131 comes almost entirely from fold 2. Fold 2 shows a drop in precision (0.479, against 0.859 to 0.999 in the other folds), while recall stays at 0.816. The cause is a single Emotet capture that contributes 1,626 flows (45\% of the fold) with no labeled C2 flows. We analyze it in Section V-D.\\

\noindent \textbf{The unit of analysis.} A flow is labeled by its destination, and
the nine beacon features are computed per endpoint triple. Flows that
share a destination are therefore not independent observations. Table
\ref{tab:unit} illustrates this effect.

\begin{table}[!t]\centering\caption{Class Balance by Unit of Analysis}\label{tab:unit}
\begin{tabular}{@{}lrr@{}}\toprule
\textbf{Unit} & \textbf{Count} & \textbf{Positive rate} \\ \midrule
Flow & 17,577 & 55.1\% \\
Endpoint triple (src, dst, port, capture) & 2,132 & 4.2\% \\
Destination address & 1,216 & 6.7\% \\
\bottomrule\end{tabular}\end{table}

No triple and no destination carries a mixed label. The 55.1\% positive rate, often cited as evidence of a balanced problem, is instead an artifact of counting flows rather than endpoints. At the endpoint level, the task is strongly imbalanced (4.2\%), which is closer to the real base rate. Three consequences follow: standard deviations at the flow level are too small; the common argument that ``the classes are balanced'' rests on an artifact; and a SOC, which blocks by destination rather than by flow, works closer to the 4.2\% regime.\\

\noindent \textbf{Feature attribution and the JA3S shortcut.} Random Forest gives
its highest weight in View B to JA3S (0.349), yet B-RF is the weakest
configuration. Many of the roughly 125 distinct JA3S values are specific
to one campaign. These values are easily separable inside a fold but useless on a held-out capture. This is campaign memorization, not learning at the
protocol level.

\subsection{Complementarity: Present but Bounded}

We score an oracle combiner as correct whenever at least one base model
is correct. It outperforms the best single view for every pair of models, so
complementarity exists. However, the room for improvement shrinks as the stronger view improves. It falls from +0.112 for the two weakest models to +0.011 for the strongest flow model paired with the TLS model. The binding limit is the set of flows that both models misclassify. This
set contains 2,545 to 2,922 flows (14\% to 17\% of the corpus), and no combiner over these base models can recover them. The two strongest models have a Q-statistic of 0.983, so they almost always make errors on the same flows, which explains the small fusion gain. Averaging four
base models reaches 0.878, matching the best combiner and capturing most of the complementarity: 0.878 against an oracle ceiling of 0.887. The rest is constrained by a block of double faults (15\%) that no combiner can reach.

\subsection{Beacon Enrichment and a Feature-Interaction Finding}

The nine beacon descriptors move the Random Forest flow view from 0.747
to 0.856, a gain of 0.109; for XGBoost, the effect is small (+0.013). We
then removed the single highest-weighted feature,
\texttt{beacon\_n\_flows} (the number of flows per endpoint triple), and
observed an unexpected result.

The two algorithms react in opposite ways. For Random Forest, the nine
features together (+0.109) provide more than the sum of their individual
contributions (+0.081). For XGBoost, the relation is reversed:
\texttt{beacon\_n\_flows} alone (0.792) outperforms all nine together (0.775).
The sign of this interaction is the reliable result. Its size (+0.028
for Random Forest, $-$0.027 for XGBoost) is below what $n = 5$ can resolve, so we report it only as a description.\\

\noindent \textbf{The feature \texttt{beacon\_n\_flows}.} This feature partly reflects a collection artifact. The strength of this feature comes partly from properties of the capture, not from real beacon behavior. Its flow count per C2 destination correlates with the capture size (Spearman $\rho = 0.718, n = 42$). Its single-feature ROC-AUC varies from 0.977 to 0.173 across folds. Furthermore, a single-threshold rule on it requires an unstable threshold (37 to 127 across folds). We report this dependence instead of removing the feature. Despite this, removing it still leaves F1 at 0.803, well above the 0.747 without any beacon features, so the beacon conclusion does not reduce to a single artifact.

\subsection{Incomplete Ground Truth}

This is the most important data finding in the study. Of the 62
captures, 20 contain no TLS flows to any indicator address, although
every capture is labeled ``Cobalt Strike.'' Direct cross-checking found
zero cases where an indicator address appears as a flow destination but
is not labeled C2, so this is \emph{not} a labeling error in our code.
Of the 20 captures, none result from an implementation bug: 16 have indicators 
but no TLS flows reach them, 3 have an empty indicator set, and 1 has
incomplete metadata. The cause is in the source data. For these
captures, the published C2 addresses never appeared as TLS-flow
destinations. The likely reasons are a non-TLS C2 channel, C2 traffic in
a separate PCAP, or indicators that describe the delivery stage rather
than the C2 stage.\\

\noindent \textbf{Bounding a related source of label noise.} One verifiable source of label noise is that 13 destinations labeled as benign share a /24 block with a known C2
indicator (1,960 flows, 11.2\% of the corpus). Sharing a /24 is not
proof of an incorrect label because hosting providers share address ranges. However, it provides an upper bound on this source. The resulting label noise
may be as large as the differences between configurations, and this
should guide how the clean-condition ranking is interpreted.

\subsection{Dual-Surface Evasion}

We applied symmetric evasion that modified all of the attacked features,
with a cluster bootstrap (2,000 resamples over the 62 captures). Each single view was unaffected by attacks on the surface it does not observe, but collapses when its own surface is forged. View A fell from 0.856
to 0.077 under a flow-side attack, which matched how View B fell under
a TLS-side attack. Therefore, the assumption that flow behavior cannot be forged does not hold. Fusion retained some ability under an attack on
either single surface. However, under a simultaneous attack on both surfaces, all
three configurations fell to approximately 0.07. This is below the trivial
always-positive baseline of 0.711. Robustness is therefore conditional on the threat model. Fusion increases the attacker's effort from forging 15 features to
54, but it gives no immunity. Furthermore, the ranking under clean conditions
does not predict the ranking under attack.

\subsection{Operational Threshold Behavior}

At deployment, the C2 base rate is significantly lower than 55.1\%, and even a
false-positive rate of 1\% would produce thousands of false alerts per day.
At that rate, early fusion recovered 13.2\% of C2 flows, while late
averaging recovered none. The gap between the laboratory F1 of 0.856 to
0.878 and this operating point is the most important practical finding in
the study. The models therefore support conclusions regarding methodology rather than deployment readiness. At these operating points, a metadata classifier is better used as a first-stage filter that narrows traffic for analyst review or SIEM correlation, rather than as a standalone blocking rule.

\section{Discussion}
\noindent \textbf{Evaluation design is a first-class result.} The largest effect
in this study is not a gain in detection, but rather a measurement artifact. One misplaced encoding step raised the F1 by 0.28, while the largest real
fusion effect was 0.022. Each of the three bias mechanisms was larger than the effects around it: leakage (0.28 F1), the unit-of-analysis shift (55.1\% to 4.2\%), and the 32\% of captures contributing one class. The practical lesson is a short checklist: report metrics at both the flow and entity level, state how many captures contribute only one class, include a trivial baseline, and measure leakage with a controlled violation.\\

\noindent \textbf{Feature quality matters more than the combination method, and
attribution is not transferable.} The single largest change in performance
is the beacon ablation (+0.109 for Random Forest), an order of magnitude
larger than any gain from a better combiner. High attribution,
however, is not evidence of a signal that transfers. JA3S has the
highest attribution in the whole study (0.349) but belongs to the
weakest configuration, and \texttt{beacon\_n\_flows} partly reflects
capture length. Attribution measures reliance inside the training fold; only a grouped or time-based split tests whether it transfers. The value of multi-view
detection for defense is therefore cost, not immunity: it forces the
attacker to forge several independently maintained surfaces at once. This constitutes defense in depth. The overhead is modest: feature extraction dominates, and tree-ensemble inference adds negligible cost, so fusion only adds a second extraction pass.

\section{Threats to Validity}

\noindent \textbf{Internal and statistical.} Grouping by capture removes the main
leakage channel. The three learned transforms were fitted inside the training fold, and a controlled violation demonstrated the impact of this constraint (0.28 F1). We selected the tree depth on the same five outer folds used for reporting.
To bound the resulting optimism, we re-ran the whole study with nested
cross-validation. This left A-RF and C-RF unchanged: the one-standard-error rule selects depth 3 in every outer fold. It also kept the ranking C-RF $>$ A-RF. Only the evasion analysis uses the cluster bootstrap that respects this structure. With $n = 5$, the smallest Wilcoxon $p$-value is 0.0625, so all clean-condition comparisons are purely descriptive. The evasion result ($p < 0.001$) reflected a large effect ($\Delta = 0.590$), not a large sample.\\

\noindent \textbf{Construct and external.} The 55.1\% positive rate at the flow
level reflects captures that focus on infections, so precision and F1
are optimistic. We reported ROC-AUC, PR-AUC, and operating points at low false-positive rates as measures that do not depend on prevalence. The
benign class is non-indicator traffic from the same sandboxed captures,
not real enterprise background traffic. With only 51.9\% of indicators
confirmed and 32\% of captures contributing a single class, the
systematic label noise may be as large as the differences between
configurations. The results describe one source, one C2 framework, and a
small set of loader families, and do not support generalization to
Sliver, Mythic, Havoc, or Brute Ratel, whose callback patterns differ.
The corpus is also skewed in time (74.2\% from 2022) and predates the
move to JA4+. These pitfalls would manifest differently elsewhere: Sliver's different TLS stack would produce non-distinctive JA3S values, different callback patterns in Mythic or Havoc would weaken the beacon features, and JA4+ would require re-examining the JA3S shortcut.

\section{Conclusion and Future Work}

We evaluated multi-view fusion for encrypted C2 detection under a leakage-controlled protocol, and found that questions about the method matter more than questions about detection. A misplaced preprocessing step raises F1 by 0.28; class balance is an artifact of counting flows rather than endpoints (55.1\% becomes 4.2\%); fusion outperforms the best single view by only 0.022; and under a simultaneous attack on both surfaces, every configuration falls below a trivial baseline.

Multi-view detection increases the attacker's cost but provides no immunity. Where a misplaced step raises F1 by 0.28 while the largest real fusion effect is 0.022, the evaluation protocol is not merely a step before the result. It \emph{is} the result.\\

\noindent \textbf{Future work.} Future directions include a time-based and leave-one-family-out split, an independent benign corpus, problem-space evasion, and a JA4+ re-evaluation.

\section*{Acknowledgment}
Khuong Nguyen-An acknowledges the support of Ho Chi Minh City University of Technology (HCMUT), VNU-HCM, for his study.

\bibliographystyle{IEEEtran}
\bibliography{references}

\end{document}